\documentclass[preprint]{aastex631}
\usepackage{amsmath}
\usepackage{soul}

\shorttitle{Efficient BWM for PTAs}
\shortauthors{J.~Sun, et al.}
\graphicspath{{./}{figures/}}
\defcitealias{aggarwal_nanograv_2020}{NG11-bwm}
\defcitealias{arzoumanian_nanograv_2015}{NG5-bwm}

\begin{document}

\title{Implementation of an efficient Bayesian search for gravitational wave bursts with memory in pulsar timing array data}

\author[0000-0002-7933-493X]{Jerry Sun}
\affiliation{Department of Physics, Oregon State University, Corvallis, OR 97331, USA}

\author[0000-0003-2745-753X]{Paul T.~Baker}
\affiliation{Department of Physics and Astronomy, Widener University, Chester, PA, 19013, USA}

\author[0000-0002-7445-8423]{Aaron D.~Johnson}
\affiliation{Center for Gravitation, Cosmology and Astrophysics, University of Wisconsin--Milwaukee, P.O. Box 413, Milwaukee WI, 53201, USA}

\author[0000-0003-2285-0404]{Dustin R.~Madison}
\affiliation{Department of Physics, University of the Pacific, 3601 Pacific Avenue, Stockton, CA 95211, USA}

\author[0000-0002-7778-2990]{Xavier Siemens}
\affiliation{Department of Physics, Oregon State University, Corvallis, OR 97331, USA}
%\affiliation{Center for Gravitation, Cosmology and Astrophysics, University of Wisconsin--Milwaukee, P.O. Box 413, Milwaukee WI, 53201, USA}



\begin{abstract}
The standard Bayesian technique for searching pulsar timing data for gravitational wave (GW) bursts with memory (BWMs) using Markov Chain Monte Carlo (MCMC) sampling is very computationally expensive to perform. In this paper, we explain the implementation of an efficient Bayesian technique for searching for BWMs. This technique makes use of the fact that the signal model for Earth-term BWMs (BWMs passing over the Earth) is fully factorizable. We estimate that this implementation reduces the computational complexity by a factor of 100. We also demonstrate that this technique gives upper limits consistent with published results using the standard Bayesian technique, and may be used to perform all of the same analyses that standard MCMC techniques can perform.
\end{abstract}

%% You will be asked to selected these concepts during the su\mathbfission process


\section{Introduction} \label{sec:intro}

%https://www.overleaf.com/project/615390e7fe2c897a0d4998fd

% Outline:
% - Pulsars have stable rotations. Measuring TOAs let us measure deviations due to GWs.

% - Status quo is to do MCMC Bayesian model parameter estimations.

% \hspace{4mm} - Pulsars have long baselines, and PTAs are growing

% \hspace{4mm} - Signal model for many GW signatures contain cross correlations that make analytic estimations difficult. 

% - Although previous work has been done searching for BWMs using MCMC and frequentist analyses (cite 11yr- and 5yr- papers). 

% \hspace{4mm} - Both searches uses the big matrix inversion, costly to evaluate likelihoods

% \hspace{4mm} - Furthermore, the Earth-term requires careful tracking of sky-selection effects, and results in MCMC taking a long time.

% \hspace{4mm} - \citetalias{arzoumanian_nanograv_2015} paper proposes expedited frequentist Earth-term analysis using factorized likelihood.

% \hspace{4mm} - In this paper, we use the scheme implemented in \citetalias{arzoumanian_nanograv_2015} to compute Bayesian posteriors for BWM model parameters. 

Millisecond pulsars (MSPs) have very stable rotations. Because the rotation is so stable, it is possible to detect small deviations in times-of-arrival (TOAs) of radio pulses from an array of these pulsars caused by gravitational waves (GWs) passing between the pulsar and radio observatories on Earth \citep{hellings_upper_1983, foster_constructing_1990, manchester_international_2013, mclaughlin_north_2013}. Pulsar timing arrays (PTAs) are expected to be able to use TOA data from many MSPs to either detect or provide constraints on GWs \citep{sazhin_opportunities_1978,detweiler_pulsar_1979}.

One signal of interest is a gravitational wave burst with memory (GW BWM). ``Memory" is a permanent change in the spacetime metric that remains after a GW passes through a region of space arising from the nonlinearity of Einstein's field equations \citep{christodoulou_nonlinear_1991,thorne_gravitational-wave_1992}. In particular, it is expected that mergers of super massive black hole binaries (SMBHBs) will leave behind detectable memory. Detections of (or constraints on) the rates of BWM events would allow for a better understanding of the rates at which these events occur in the universe \citep[e.g.][]{islo_prospects_2019}. Additionally, because all GW events leave behind GW memory, detections of BWMs could lead to discoveries of new sources of GWs \citep{cutler_gravitational-wave_2014}.

A GW BWM passing over an Earth-pulsar pair will shift the pulsar's observed rotational frequency \citep[e.g.,][]{van_haasteren_gravitational-wave_2010}. This shift causes a difference between the observed frequency of the pulsar and the frequency expected from the timing-model fit, and will therefore contribute to potentially detectable signal in the pulsar's TOAs \citep{seto_search_2009, pshirkov_observing_2010, van_haasteren_gravitational-wave_2010, cordes_detecting_2012, madison_assessing_2014, islo_prospects_2019}. The observed rotational frequency may change to be either faster or slower depending on the orientation and polarization of the memory wavefront, which determines the sign of the memory.  

In this paper, we will discuss the adaptation of analysis techniques used in the NANOGrav five-year search for GW BWMs \citep[][hereafter \citetalias{arzoumanian_nanograv_2015}]{arzoumanian_nanograv_2015} to expedite the Bayesian methods used in the NANOGrav 11-year search for BWMs  \citep[][hereafter \citetalias{aggarwal_nanograv_2020}]{aggarwal_nanograv_2020}. In \citetalias{aggarwal_nanograv_2020}, no detection of GW BWMs was reported. Thus, the authors presented Earth-term upper limits as a function of burst epoch and sky location (among other results, but these will be our focus). Our goal in this paper is to show that the adapted techniques from \citetalias{arzoumanian_nanograv_2015} may be used to efficiently perform Bayesian analyses comparable to those in \citetalias{aggarwal_nanograv_2020} with a similar degree of accuracy.

In \autoref{sec:signalmodel}, we describe the effect of a BWM on the TOA residuals of a pulsar. In \autoref{sec:prev_methods}, we discuss the current standard Bayesian approach to searching for GW BWMs in PTA data and the implementation of an efficient technique for speeding up this search. In \autoref{sec:results}, we compare the results of upper-limit calculations using our more efficient technique against results previously published in the literature. We also discuss the improvements in computational efficiency that come from this technique.

\section{Signal and Data Model}\label{sec:signalmodel}

The rise time for the memory component of a GW BWM is much shorter than the typical observing cadence of PTAs; thus we may ignore it and consider the frequency-shifting effect to be instantaneous \citep{favata_gravitational-wave_2010, van_haasteren_gravitational-wave_2010, madison_assessing_2014}. This manifests as a linear ``ramp" in the residuals, since a constant excess or deficit of pulse phase will accrue with each rotation of the pulsar. Consider a memory event from a source propagating in the direction $\mathbf{\hat{k}}$, with strain $h_{mem}$, passing over the Earth from which we observe a pulsar located in the direction $\mathbf{\hat{p}}$. The memory wavefront has a polarization angle described by an angle $\psi$ (where $\psi$ is the angle between the principal polarization vector of the wave and the projection of the line-of-sight to the pulsar on the plane perpendicular to the propagation direction of the wave). Following \citetalias{arzoumanian_nanograv_2015} and \citetalias{aggarwal_nanograv_2020}, the perturbation to pulse times of arrival, $\delta t$, may be modeled as:

\begin{equation}\label{eqn:bwm_on_resids}
\delta t_{\text{bwm}}(t) = B(\mathbf{\hat{k}}, \mathbf{\hat{p}}, \psi) \vspace{2mm} h_{\text{mem}}(t),
\end{equation}
\noindent
where $h_{\text{mem}}(t)$ is the time-dependent strain of the memory wave front, and the geometric factor $B$ accounts for the relative orientation of the source and pulsar \citep{estabrook_response_1975, hellings_upper_1983}. The geometric projection factor is:
\begin{equation} \label{eqn:geometric_factor}
    B(\mathbf{\hat{k}}, \mathbf{\hat{p}}, \psi) = \frac{1}{2} \cos(2 \psi)(1 - \cos{\alpha}),
\end{equation}
\noindent
where $\alpha$ is the angle between $\mathbf{\hat{p}}$ and $\mathbf{\hat{k}}$ (pulsar location and propagation direction, respectively) .

The time-dependent strain term is \citep {van_haasteren_gravitational-wave_2010, pshirkov_observing_2010}:
\begin{equation}\label{eqn:bwm_wavefront_strain}
h_{\text{mem}}(t) = h_0 \vspace{4mm} [(t - t_0) \Theta(t - t_0) - (t - t_i)\Theta(t - t_i)]
\end{equation}

where $h_0$ is the strain of the memory; $t_0$ is the time that the memory wave front passed over the earth; $t_i = t_0 +(|\vec{\mathbf{p}_i}|/c) \hspace{1mm} [1 + \cos(\theta_i)]$, the time at which the memory wave front passed over the $i$-th pulsar. Because each pulsar in NANOGrav's 11-year data release, and more generally, current PTA datasets, is on the order of thousands of light-years away from the Earth, and total observation times are order tens of years, we only expect that one of the two terms in \autoref{eqn:bwm_wavefront_strain} will be nonzero. The first term in \autoref{eqn:bwm_wavefront_strain} is called the ``Earth term", and the second is called the ``pulsar term". A BWM may be observed either when it passes over a single pulsar, or when it passes over the Earth. In the former case, we will see the frequency of a single pulsar spontaneously change. In the latter case, we expect to see the rotational frequency of each pulsar change simultaneously with a characteristic quadrupolar amplitude pattern. In either case, the time at which the BWM wave front causes an apparent rotational frequency change is defined as the burst epoch.

\section{Methodology}\label{sec:prev_methods}

We begin by discussing the standard Bayesian approach to searching for a GW BWM. This discussion will summarize the approach taken in \citetalias{aggarwal_nanograv_2020}, which searched the NANOGrav 11-year data set for GW BWMs. Then, we will discuss the the adaptation of the techniques used in \citetalias{arzoumanian_nanograv_2015} that expedites both the pulsar- and Earth-term searches.

\subsection{Bayesian Approach}\label{sub:bayesian}

\citetalias{aggarwal_nanograv_2020} modeled the timing residuals $\mathbf{\delta t}$ of a pulsar as: 
\begin{equation}\label{eqn:residual_model}
    \mathbf{\delta t} = \mathbf{\delta t}_{bwm} + M\boldsymbol\epsilon + F\mathbf{a} + \mathbf{n},
\end{equation}
where $\mathbf{\delta t}$ are the remaining perturbations to the TOAs from a pulsar after fitting parameters in the pulsar's timing model using a general-least-squares fit \citep{arzoumanian_nanograv_2018}. These remaining perturbations, the timing residuals, are expected to originate from a combination of noise processes, errors in the timing model fit, and GW signals. $\mathbf{\delta t}_{bwm}$ are the contributions to the timing residuals from a gravitational wave BWM. $M$ is the design matrix of the linearized timing model which accounts for uncertainty in the residuals from the imperfect timing model fit $\boldsymbol\epsilon$. The elements of vector $\mathbf{n}$ are Gaussian white noise uncertainties in the observed TOAs. Finally, $F$ is the design matrix for pulsar intrinsic red-noise, modeled as a Fourier series with coefficients $\mathbf{a}$.

The red-noise spectrum, for example from a stochastic background of GWs, is expected to behave as a power law \citep{phinney_practical_2001}:
\begin{equation}\label{eqn:rn_plaw}
    P(f_j) = A_{j}^2\left(  \frac{f_j}{\text{yr}^{-1}} \right)^{-\gamma},
\end{equation}
where $P(f_j)$ is the power spectral density of the red-noise process and $A_j$ is the characteristic amplitude of the red-noise process in the $j$-th frequency bin using a reference frequency of $\text{yr}^{-1}$.

From \autoref{eqn:residual_model}, we can construct an approximation of the Gaussian white noise given an estimation of the model parameters:
\begin{equation}\label{eqn:wn_resids}
    \mathbf{n} = \mathbf{\delta t} - \mathbf{\delta t}_{\text{bwm}} - M\boldsymbol\epsilon - F\mathbf{a} .
\end{equation}
This is only an approximation of the white noise since the terms on the right-hand side are estimations. However, if the white noise is expected to be Gaussian, we can write the probability of observing this particular series of white noise residuals as:
\begin{equation}
    p(\mathbf{n})=
    \frac{
        \exp{(-\frac{1}{2}
            \mathbf{n}^T N^{-1}\mathbf{n}   
        )}
    }{
        \sqrt{2 \pi \det{N}}
    }
\end{equation}

where $\mathbf{N}$ is a covariance matrix of white-noise uncertainties in each observed TOA.

Then, the likelihood of a BWM signal in the pulsar timing residuals is equivalent to the likelihood that the remaining residuals after subtracting out deterministic effects is Gaussian white noise. In other words:
\begin{align}\label{eqn:likelihood}
\begin{split}
    p( \mathbf{\delta t} | \boldsymbol\epsilon, \mathbf{a}, \mathbf{\delta t}_{\text{bwm}}) &= p( \mathbf{\delta t} | \boldsymbol\epsilon, \mathbf{a}, h_{\text{mem}}, t_0, \mathbf{k},\mathbf{p}, \psi) \\
     &= \frac{\exp{[-\frac{1}{2}( \mathbf{\delta t} - \mathbf{\delta t}_{\text{bwm}} - M\boldsymbol\epsilon - F\mathbf{a})^T N^{-1}( \mathbf{\delta t} - \mathbf{\delta t}_{\text{bwm}} - M\boldsymbol\epsilon - F\mathbf{a} )]}}{\sqrt{2 \pi \det{N}}},
\end{split}
\end{align}

where above, we have explicitly written out the parameters that determine $\mathbf{\delta t}_{\text{bwm}}$. This parameter space, when including each Fourier coefficient and timing model parameter, is very high-dimensional. 

It is possible to analytically marginalize the likelihood in \autoref{eqn:likelihood} over the parameters that describe the Gaussian processes and reduce the dimensionality of the parameter space \citep{lentati_hyper-efficient_2013,van_haasteren_new_2014, van_haasteren_low-rank_2015}. The reduced likelihood is:
\begin{equation}\label{eqn:marginalized_likelihood}
p( \mathbf{\delta t} | h_{\text{mem}}, t_0, \mathbf{k},\mathbf{n}, \psi) = \frac{\exp{(-\frac{1}{2} \mathbf{q}^T C^{-1} \mathbf{q})}}{\sqrt{2 \pi \det{C}}}
\end{equation}
where
\begin{align}
    \mathbf{q} =  \mathbf{\delta t} -  \mathbf{\delta t}_{\text{bwm}}
\end{align}
is the residual due to the deterministic BWM only, and
\begin{align}
    C = N + TDT^T
\end{align}
is a noise covariance matrix accounting for white noise, $N$, and .
$$
T = 
    \begin{bmatrix}
    M & F
    \end{bmatrix}, \quad
    \mathbf{b} =
    \begin{bmatrix}
    \boldsymbol\epsilon \\
    \mathbf{a}
    \end{bmatrix}, \quad
    D =
    \begin{bmatrix}
    M & 0\\
    0 & F
    \end{bmatrix}.
$$

where $D_{\text{tm}}$ and $D_{\text{rn}}$ are Gaussian priors on the the timing model parameters and red noise parameters, respectively.

This likelihood is implemented in the  \textsc{ENTERPRISE} pulsar-timing GW analysis software package \citep{ellis_justin_a_enterprise_2020}. 

Now that the likelihood has been constructed, samples from the posterior distributions are drawn using the Markov-Chain Monte Carlo sampler implemented in the \hbox{\textsc{PTMCMCSampler}} package \citep{ellis_jellis18ptmcmcsampler_2017}. 

Great care must be taken when computing upper limits over the sky because of a strong selection bias. If there is no support for a signal in the data, then the maximum posterior probability will be determined largely by the prior. Because our amplitude prior spans many orders of magnitude, there is much more prior volume at higher amplitudes. This means the posterior will be maximized for bursts with very large amplitudes at insensitive areas of the sky. Because the burst is placed at an insensitive area of the sky, the data cannot exclude this strong signal. This will cause the 1D marginal posterior to be biased towards very high amplitude for combinations of burst epochs, sky locations, and polarization angles where the PTA has low sensitivity. This would not fairly represent the sensitivity of the Earth-term search \citepalias{aggarwal_nanograv_2020}.

To remedy this, \citetalias{aggarwal_nanograv_2020} sampled individual ``source-orientation" bins, in which the burst epoch, sky location, and polarization are all fixed. Then, a full Earth-term posterior is constructed by concatenating an equal number of samples from each source-orientation bin. This sampling scheme is the equivalent of implementing a prior which exactly cancels the selection effect, resulting in a posterior that is uniform in source-orientation. This is related to the technique used in \citet{malmquist_relations_1922}.

 More specifically, to place an amplitude upper limit as a function of burst epochs, \citetalias{aggarwal_nanograv_2020} used 48 \textsc{HEALPix} sky bins, with 8 polarizations in each sky bin. This gives a total of 384 source-orientation bins in each of the 40 burst epoch bins. An MCMC sampler is then used to sample the posterior probability distributions of the BWM amplitude. Then, to compute an amplitude posterior marginalized over source-orientations for a fixed burst epoch, equal numbers of samples are taken from each source-orientation bin and concatenated. 
 
 To place upper limits as a function of sky position, \citetalias{aggarwal_nanograv_2020} used 768 \textsc{HEALPix} sky bins and directly sampled polarization (rather than sampling in fixed polarization bins). Then, the amplitude upper limit may be computed from the marginalized amplitude posterior for each sky position.

For a summary of the priors, see Table \ref{tab:prior_distributions}.

\begin{deluxetable}{ccc}
\tabletypesize{\footnotesize}
\tablewidth{0pt}
\tablecolumns{3}
\tablehead{\colhead{Parameter} & \colhead{Prior} & \colhead{Description}}
\startdata
$\log_{10}A_{\text{RN}}$ & $\text{LinearExp}(-17, -11)$ & Amplitude of intrinsic pulsar red noise \\
$\gamma_{\text{RN}}$ & $\text{uniform}(0, 7)$ & Spectral index of intrinsic pulsar red noise \\
$\log_{10}A_{\text{BWM}}$ & $\text{LinearExp}(-17, -10)$ & Amplitude of global BWM \\
$\psi_{\text{BWM}}$ & $\text{Uniform}(0, \pi)$ & Polarization of BWM \\
$\theta_{\text{BWM}}$ & $\text{Uniform}(0, \pi)$ & Polar angle of BWM source \\
$\phi_{\text{BWM}}$ & $\text{Uniform}(0, 2\pi)$ & Azimuthal angle of BWM source \vspace{3mm} \\
 $t_{\text{BWM}}$ & 
\begin{tabular}{cc}
$\text{Uniform}(\text{MJD }56000, \text{MJD }57000)$ \\
$\text{Uniform}(\text{MJD }53216, \text{MJD }57387)$ 
\end{tabular} & Global BWM epoch for UL vs. skyposition (upper) and UL vs. epoch (lower)
\enddata
\caption{\label{tab:prior_distributions} Priors used for each of the model parameters in the Bayesian search for global Earth-term GW BWMs using the full PTA. There are a total of five global BWM parameters, as well as two parameters for each pulsar in the PTA. The priors on the logarithm of the amplitude are equivalent to setting uniform priors over the amplitude. Because of selection effects, it is nontrivial to implement uniform priors over the sky location of the burst. More details on this may be found in \autoref{sub:bayesian}. The prior on $t_{\text{BWM}}$ also varies depending on the particular upper limit calculation. For upper limits as a function of sky location, we use priors between MJD 56000 and MJD 57000. For upper limits as a function of burst epoch, we use priors that encompass all the timing data (approximately MJD 53216 to MJD 57387). There is more detail on the burst epoch prior in \autoref{sub:accelerated}.}
\end{deluxetable}

\subsection{Accelerated Bayesian Search}\label{sub:accelerated}

For the accelerated Bayesian search, we mimic the Bayesian approach described in \autoref{sub:bayesian} as closely as possible. We found that the computational cost of the MCMC sampling required was prohibitively expensive to perform on machines we have access to. Thus, to expedite the Bayesian search, we leverage a fact from \citetalias{arzoumanian_nanograv_2015}: the Earth-term likelihood is able to be factorized into a product of pulsar-term likelihoods. In other words:
\begin{align}\label{eqn:factlike}
\begin{split}
    p(\mathbf{\delta t} | \hat{\mathbf{k}}, \psi_k, t_B, h_B) &= \prod_{K=1}^{M}{p_{K}(\mathbf{\delta t} | \hat{\mathbf{k}}, \psi_k, t_B, h_B}) \\
    &= \prod_{K=1}^{M}{p_K(\mathbf{\delta t}|h_{K}, t_B)},
\end{split}
\end{align}
and:
\begin{equation} \label{eqn:factorized_geom_factor}
    h_K =  B(\hat{\mathbf{k}}, \hat{\mathbf{p}}_K, \psi) \times h_B,
\end{equation}
where $p(\mathbf{\delta t} | \hat{\mathbf{k}}, \psi_k, t_B, h_B)$ is the global likelihood of a burst with sky location $\hat{\mathbf{k}}$, polarization $\psi_k$, epoch $t_B$, and characteristic strain $h_B$. $p_K$ is the pulsar-term likelihood of this burst in the $K$-th pulsar, with $h_K$ being the observed amplitude of the burst after accounting for the geometric projection, $B(\hat{\mathbf{k}}, \hat{\mathbf{p}}_K, \psi)$, of the burst onto the pulsar-Earth line-of-sight. As pointed out in \citetalias{arzoumanian_nanograv_2015}, the pulsar's TOAs have no information about the parameters of the burst other than the apparent burst amplitude after being projected onto the pulsar-Earth line-of-sight. This allows us to pre-compute the individual pulsar-term likelihoods over a grid of only post-projection BWM amplitude and burst epoch without losing any information. Then at run time, the geometric projection factor, \autoref{eqn:geometric_factor}, may be applied to give the correct post-projection amplitude for any given global trial burst. This way, we may then look up the corresponding likelihoods of the global burst in the pre-computed lookup tables and combine them using \autoref{eqn:factlike}.

With this in mind, we begin the accelerated Bayesian search by first generating five-dimensional lookup tables for the likelihood of each pulsar (the far right-hand side of \autoref{eqn:factlike}). In addition to the BWM amplitude $|h_K|$, epoch $t_B$, and the sign of $h_K$, we include the amplitude $A_{rn}$ and spectral index $\gamma$ of a red-noise process described in \autoref{eqn:rn_plaw}. We emphasize that we must keep track of the sign separately because any trial BWM may delay or advance the TOAs depending on the relative orientation of the BWM polarization to the pulsar-Earth line of sight. Recall that a single BWM has a quadrupolar antenna pattern. Consider a pulsar that is in a part of the sky such that a particular BWM would cause the TOAs to be advanced by some amount. Rotating the trial burst by 90 degrees would cause the TOAs to be delayed---rather than advanced---by the same amount. Thus, for global searches of BWMs with every possible orientation, we have to include the likelihoods for every amplitude of BWM with both positive and negative signs. We then numerically integrate over the red-noise parameters using composite Simpson's rule to obtain one red-noise-marginalized three-dimensional likelihood lookup table for each pulsar (with the remaining parameters being $\{|h_K|, t_{B}, \text{sign}(h_K) \}$). Then, we may compute marginal amplitude likelihoods for any pulsar-term BWM by integrating over the burst epoch. 

Next, we can combine the pulsar-term likelihood tables to construct global likelihood lookup tables which contain the Bayesian likelihoods of finding an Earth-term trial burst with fixed sky position $\Omega$, polarization $\psi$, and strain $h_{mem}$ at some fixed trial burst epoch $t_0$. To do so, we project a burst with these fixed global parameters onto each pulsar line-of-sight to find the amplitude and sign with which this burst will appear in the pulsar's timing residuals. This allows us to compute the observed amplitude in each of the pulsar terms. We then simply look up the likelihood for each pulsar-specific projected amplitude in the single-pulsar lookup tables. Finally, the global likelihood of this trial burst is computed by multiplying the pulsar-term likelihoods, \autoref{eqn:factlike}. We compute one two-dimensional lookup table varying over trial bursts characterized by $(h_{mem}, t_0)$ for each set of trial parameters $(\Omega, \psi)$.

We can then construct global amplitude posteriors as a function of sky position and epoch by integrating out any nuisance parameters against their prior distributions. We do so in the same way as in \citetalias{aggarwal_nanograv_2020} (described in \autoref{sub:bayesian}); whenever we marginalize over source-orientation, we are careful to do this by taking equal samples from each source-orientation bin to demand a posterior that is uniform in source orientation.

Specifically, to compute the upper limits as a function of burst epoch, we compute two-dimensional posterior distributions of BWM amplitude and epoch in each of 48 \texttt{HEALPix} sky pixels with one of eight fixed polarizations (for a total of 384 total source-orientation bins). Then, we compute the marginal BWM amplitude posterior for each trial burst epoch. concatenate samples from each source-orientation bin to compute the full-sky, polarization-marginalized 95\% upper limits as a function of burst epoch. 

To compute the upper limits as a function of sky location, we use 768 \texttt{HEALPix} skypixels and eight polarization bins, with the prior for BWM epochs limited between MJD 56000 and MJD 57000. We use this limited prior because after MJD 56000, there are no new pulsars added to the PTA. It is challenging to come up with a scheme for determining representative BWM amplitude posteriors over a period in which new pulsars are continually added, so we limit our search only to the period in which we already have data for each pulsar. For each source-orientation bin, we then marginalize over burst epochs to obtain the marginal BWM amplitude posterior, and concatenate samples from all eight polarization bins. Finally, we marginalize over polarization by concatenating samples from each polarization bin to obtain the marginal amplitude posterior for each sky location. In very brief summary:
\begin{enumerate}
    \itemsep0em
    \item Compute pulsar-term BWM likelihoods on a grid of $\{\log_{10}{|h_K|},\text{sign}(h_K) ,t_0, \log_{10}{A_{\text{rn}}}, \gamma_{\text{rn}}\}$
    \item Marginalize pulsar-term BWM likelihoods over red noise parameters.
    \item Use pulsar-term likelihoods, \autoref{eqn:factlike} and \autoref{eqn:factorized_geom_factor} to compute Earth-term BWM likelihoods on a grid of $\{ \log_{10}{h_B}, t_B\}$ for each set of trial burst parameters $\theta,\phi, \psi$
    \item Marginalize over: 
    \begin{enumerate}
        \itemsep0em
        \item Burst epoch and polarization to compute amplitude posterior over sky location
        \item Sky location and polarization to compute amplitude posterior over burst epoch
    \end{enumerate}
\end{enumerate}

We computed these global BWM amplitude posteriors using a prior that is log-uniform in the burst amplitude. However, to compute upper limits on the burst amplitude, we need to use a posterior with a prior that is \textit{uniform} in the burst amplitude. Although our marginal posteriors have log-uniform priors built in, we can still readjust the prior. Under a log-uniform prior, the burst amplitude posterior is:
\begin{align}
\begin{split}
    p_{\text{log-uni}}(A_{\text{bwm}}|d) &\propto p(d|A_{\text{bwm}}) \hspace{1mm} \pi_{\text{log-uni}}(A_{\text{bwm}})\\
    &\propto p(d|A_{\text{bwm}}) \hspace{1mm} \frac{1}{A_{\text{bwm}}}
\end{split}
\end{align}

where $\pi(A_{\text{bwm}}) \propto \frac{1}{A_{\text{bwm}}}$ is the prior distribution on $A_{\text{bwm}}$. We can see that multiplying the (log-uniform) posterior by the amplitude will then correctly adjust the prior to have equal volume at each burst amplitude instead of equal volumes at each order-of-magnitude of burst amplitude. Once this posterior is recomputed with the correct prior, we can compute the 95\% amplitude upper limits by numerical integration or rejection sampling.

We would also like to emphasize a new, unique advantage of this accelerated search for BWMs. One challenge of using PTAs to detect gravitational waves is the necessity of accurate, well-understood pulsar noise models. Because our global likelihood is computed using individual pulsar-term likelihoods, we are free to experiment with different noise models for each pulsar individually. In contrast, using the traditional techniques would require a full recomputation of the Bayesian posteriors using MCMC, even when altering just one of the pulsars' noise models. There has been much work done to improve pulsar noise models, and this factorized approach very robustly allows for adjustments of noise models during analysis while minimizing the computational cost. Furthermore, this would also allow us to use bespoke noise models for each pulsar if necessary.

\section{Results} \label{sec:results}
In this section, we will compare upper limits on the amplitudes of GW BWMs computed using this more efficient search and the previously published upper limits in \citetalias{aggarwal_nanograv_2020}. Then, we discuss the improvements in efficiency. 

\subsection{Pulsar-Term Comparisons}

\begin{figure}
    \centering
    \includegraphics[width=1\columnwidth]{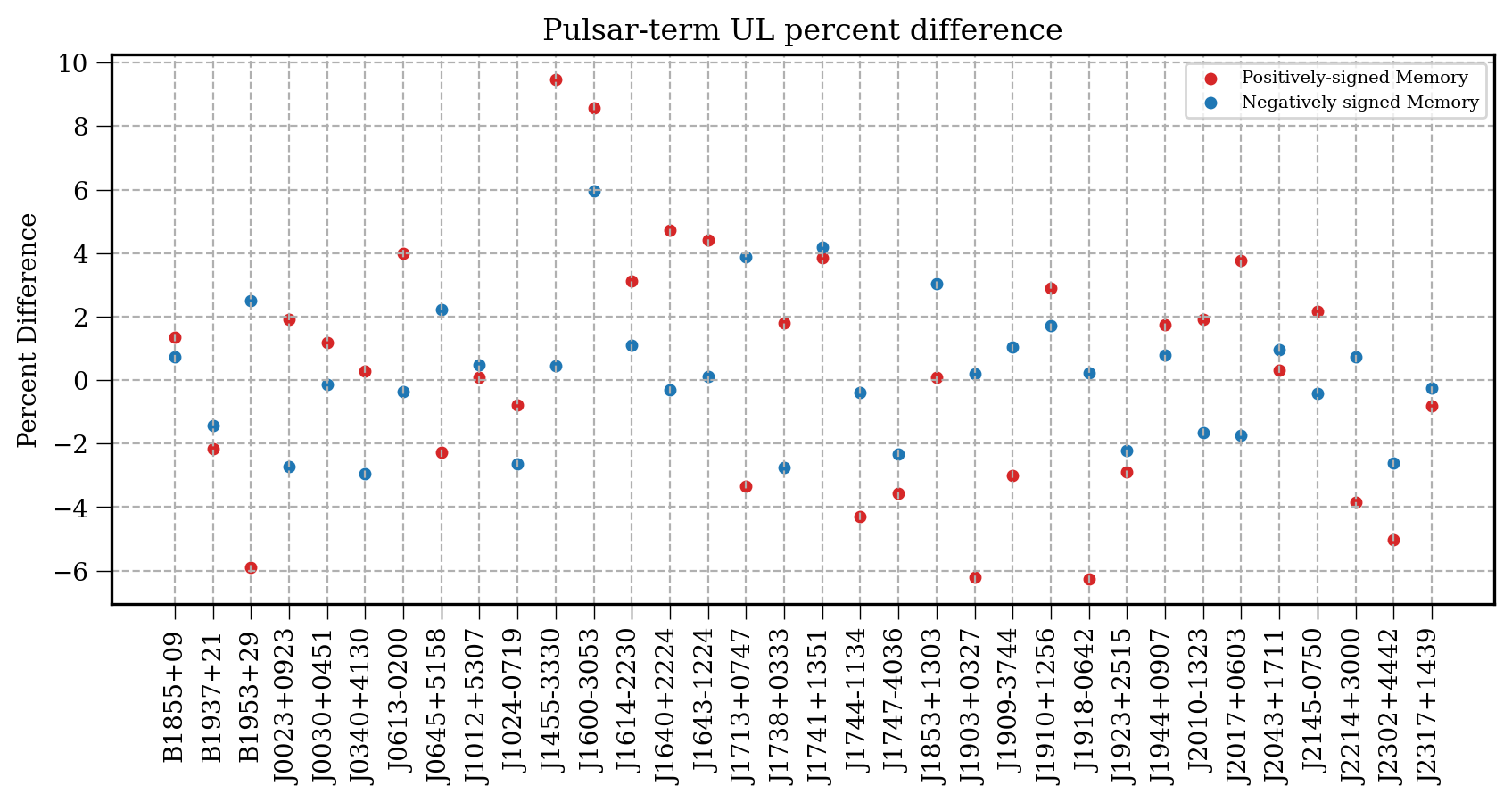}
    \caption{Percent difference in the pulsar term amplitude upper limits for each pulsar used in \citetalias{aggarwal_nanograv_2020}. We see good agreement, with percent differences less than 10\% . For this comparison, we only searched the middle 80\% of the TOA data for each pulsar. This is because many pulsars have very sparse observations early on. Furthermore, there will be little evidence for a BWM near the end of a data set since there will not be enough observed TOAs after the trial epoch to accurately detect a BWM. The red points are the percent differences for amplitude upper limits of positively-signed memory, and the blue points are the differences for negatively-signed memory.}
    \label{fig:psr_term_UL_pctdiffs}
\end{figure}

\autoref{fig:psr_term_UL_pctdiffs} shows the percent difference between the pulsar-term BWM upper limits for both positively- and negatively-signed memory computed using direct MCMC methods and our lookup-table-based method. We chose to compare the differently-signed memory upper limits separately in order to fully compare the two techniques. 
Additionally, we narrowed the priors on burst epoch to only search for bursts in the middle 80\% of the data set of each individual pulsar. This is because it is difficult to detect bursts early and late in the data set. If there is not enough data on either side of the BWM, it is possible to fit an very large bursts to the data, which will cause the amplitude upper limits to be overestimated.

%\js{I'm starting to think this isn't the best way to compare these. Should we add an associated violin plot or something of each pulsar's marginal amplitude posterior? What is the most informative way to show the pulsar-term ULs?}

We see that the new method produces amplitude upper limits within 10\% of the MCMC methods with no obvious systematic errors. We thus expect any differences in the resulting ULs to be a result of using a finite grid in the creation of the pulsar-term likelihoods lookup tables.

% \startlongtable
% \begin{deluxetable}{c}
% \tablewidth{4}
% \tablecolumns{4}
% \tablehead{
%     \colhead{heading1} & \colhead{heading2} & \colhead{heading3} & \colhead{heading4} 
% }
% \startdata
% 1 & 2 & 3 & 4 \\
% \enddata
% \end{deluxetable}

\subsection{Earth-Term Comparison}
\begin{figure}
    \centering
    \includegraphics[width=0.95\columnwidth]{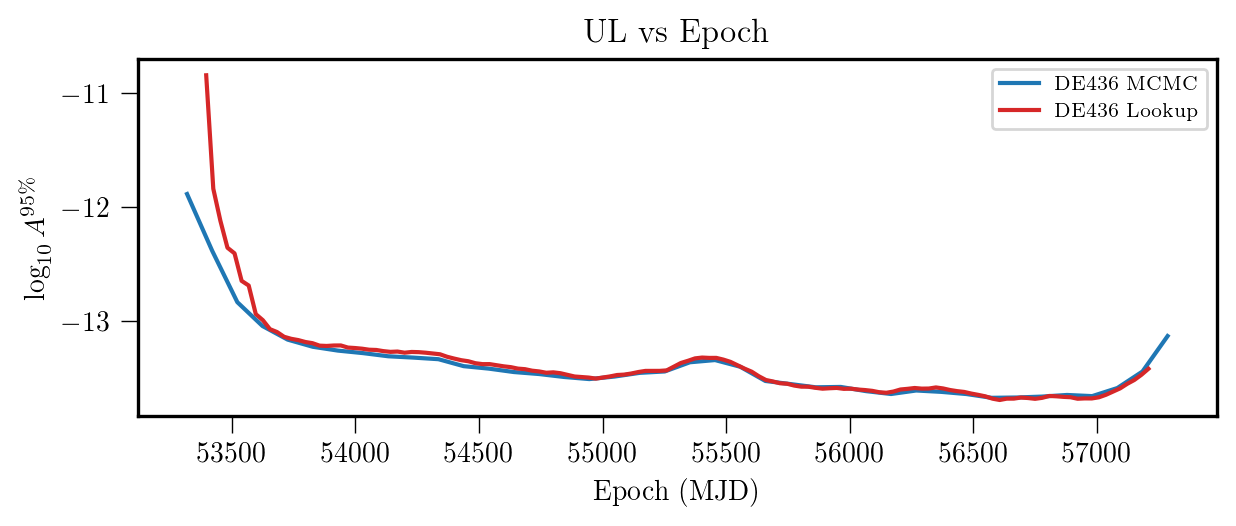}
    \caption{The 95\% BWM amplitude upper limits as a function of observation epoch. The original results are plotted in red, and the blue curve is used with permission from the authors of \citetalias{aggarwal_nanograv_2020}. There is good agreement for the vast majority of the data set, with some discrepancy at early times. We believe that these discrepancies arise from the lack of data early in the data set, and expect uninformative, unconstraining upper limits at these trial burst epochs.}
    \label{fig:UL_vs_epoch}
\end{figure}

\begin{figure}
    \centering
    \includegraphics[width=0.9\columnwidth]{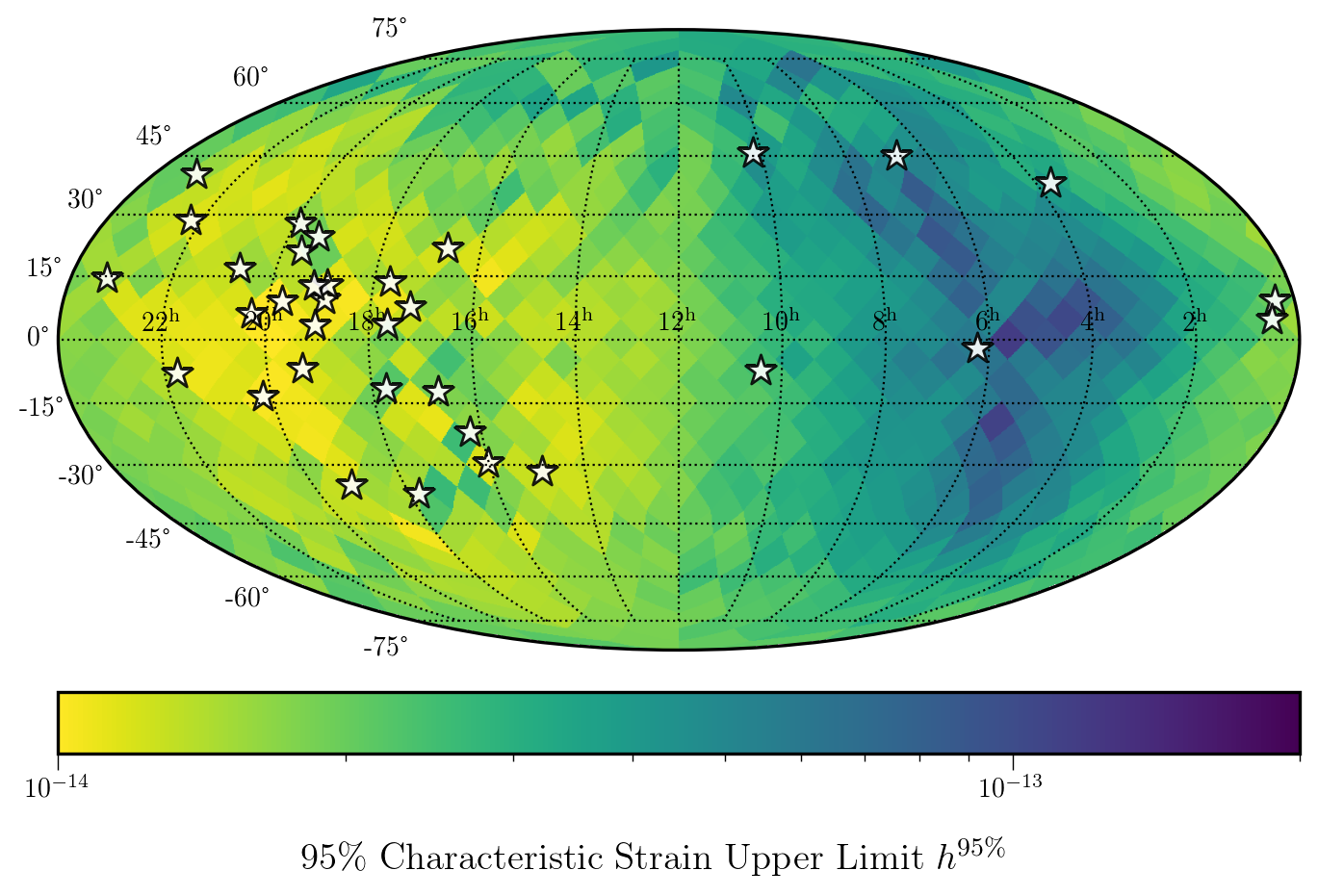}
    \caption{\textbf{Left:} 95\% BWM amplitude upper limits as a function of sky location. The stars mark the locations of the pulsars in NANOGrav's 11-year data release. As expected, the PTA is most sensitive to BWMs in sky locations where many pulsars are being timed.}
    \label{fig:UL_vs_skypos}
\end{figure}

For the Earth-term upper limits, we report two results: 1) the upper limits as a function of burst epoch and 2) the upper limits as a function of position in the sky. These results are shown in \autoref{fig:UL_vs_epoch} and \autoref{fig:UL_vs_skypos}, respectively.

In \autoref{fig:UL_vs_epoch}, we see that both methods return nearly identical upper limits as a function of burst epoch. There are some significant differences, however, at early epochs. Although the upper limits appear very discrepant, at these early epochs, there are very few recorded TOAs. As such, it is impossible to place very accurate limits on a BWM, since very large amplitude BWMs can be fit to the sparse data. Therefore, despite the apparent differences, we are not very concerned, since we expect a very non-constraining upper limit at these early epochs. More importantly, as more pulsars and more data are added to the PTA, the upper limits become nearly identical.

\autoref{fig:UL_vs_skypos} shows the upper limits on the BWM amplitude as a function of sky location using the method described in \autoref{sub:accelerated}. The resulting amplitude posterior is sampled to compute the 95\% upper limit. We find that the amplitude upper limits as a function of sky location are similar to those reported in \citetalias{aggarwal_nanograv_2020}. 

Although we can comment on general similarities between the results, we cannot directly compare them. \citetalias{aggarwal_nanograv_2020} included an additional model, called \texttt{BayesEphem}, in their analysis. This model accounts for uncertainty in the solar system ephemeris. This is especially important the NANOGrav 11yr dataset, because the observation baseline is very close to Jupiter's orbital period.

This model introduces eleven extra parameters, which is far too many to use on our parameter grid. It is therefore impossible to include \texttt{BayesEphem} using the techniques described in this work. Since there are no published results for upper limits on BWMs in the NANOGrav 11-year data set as a function of sky position that do not include \texttt{BayesEphem}, we report our results without a comparison.

This technique of using pulsar-term likelihood tables can be used to reproduce the same types of analyses and results that MCMC-based methods can. The fundamental Bayesian methodology is identical; both techniques compute marginalized posterior probabilities for model parameters. This method simply takes advantage of the factorizable likelihood to more efficiently carry out the marginalization.

\subsection{Computational Improvement}

The computational complexity of computing the pulsar-term lookup tables is dominated by the cost of inverting the covariance matrix in \autoref{eqn:marginalized_likelihood}. This is an $M\times M$ matrix, where $M$ is the number of Gaussian process parameters, needed for a single pulsar-term BWM signal model \citep{ellis_jellis18ptmcmcsampler_2017, ellis_justin_a_enterprise_2020}. One inversion has computational complexity $O(M^3)$ To compute a full pulsar-term lookup table, we evaluate the likelihood once for each point on a five-dimensional grid ($A_{\text{rn}}, \gamma_{\text{rn}}, |h_K|, \text{sign}(h_K), t_0$). Thus, the total cost of the inversions we must perform for one lookup table is $N_{A\text{rn}} N_{\gamma} N_{A\text{bwm}} N_{\text{sign}} N_{t_0} M^3$. For this paper, for a pulsar which has 10 years of data, the total number of grid points is approximately $32 \times 10^6$. Then, if we compute one lookup table for each pulsar, for the purpose of convenient comparison, we can consider the complexity to be approximately $10^7$ $N_{psr} M^3$.

The computational complexity of one pulsar-term search for BWMs using an MCMC sampler may be approximated to be just the product of the complexity of one likelihood evaluation and the number of evaluations needed. This means the complexity of the pulsar-term search for BWMs is approximately $N_{\text{iter}} N_{psr} M^3  $, where $N_{\text{iter}}$ is the number of iterations used per sampling run. Normally, $N_{\text{iter}} \approx 10^6$ is sufficient for parameter estimates to converge, so we can consider the complexity to be $10^6$ $N_{psr} M^3$.

It is very clear that the cost of producing one lookup table is significantly more expensive than performing one pulsar-term BWM search. However, once the pulsar-term likelihood tables are computed, it is very cheap to compute the global likelihoods in a full-PTA, Earth-term BWM search. For example, to compute the upper limits as a function of trial burst epoch (the results shown in \autoref{fig:UL_vs_epoch}), a full-PTA covariance matrix must be inverted. Because the signal model does not contain correlations between pulsar pairs, we may take advantage of the block diagonal structure of the covariance matrix and invert it in $O(N_{psr}  M^3)$. In other words, the matrix inversion itself is no less expensive. However, because of the sampling scheme, we must perform one MCMC sampling run for each set of $(N_\Omega, N_\psi, N_t)$. Thus, the total complexity of computing upper limits as a function of burst epoch is approximately $N_{\text{iter}} N_\Omega N_\psi N_t N_{psr} M^3$. In \citetalias{aggarwal_nanograv_2020}, this total cost is approximately $8 \times 10^8$ $N_{psr} M^3$. 

We see that, although the search is less efficient for computing pulsar-term upper limits, it is far more efficient when computing certain full-PTA searches. On an Intel i9-9900K CPU with 8 physical cores operating at 3.60GHz, it takes approximately two weeks to compute all the single pulsar lookup tables. Once the lookup tables have been produced, each of the full-PTA searches may be completed in approximately two days. Using only MCMC sampling to compute full-PTA upper limits would have taken approximately three years.

\section{Conclusion}
In this paper, we have implemented a more efficient technique for performing a Bayesian search for GW BWMs by circumventing repeated, expensive matrix inversions. While the method is faster, the upper limits computed in this paper are less constraining than previous results. This is because the global posteriors computed on a grid do not accurately characterize the high-probability regions of parameter volume as well as adaptive MCMC methods. We believe that there are ways to improve the upper limits. For example, a robust solution to help the pulsar-term upper limits converge with MCMC-derived upper limits may be implementing a scheme for adaptive grid spacing depending on the local variation of the likelihood surface. This way, we would spend less time over-characterizing regions of parameter space that do not vary much, while maintaining accuracy in quickly-varying regions of parameter space.

Overall, we find that our sky-averaged upper limits as a function of burst epochs (see \autoref{fig:UL_vs_epoch}) match well with previously published results, with almost no difference in the most sensitive regions of the data set. Although the upper limits differ somewhat significantly at early trial epochs. This is somewhat unsurprising; there is very little timing data at early epochs, and we expect very weak constraints on any BWMs appearing this early in the data set. 

Furthermore, we are able to perform the same full-PTA search for GWs over the entire sky. Although we cannot compare results with \citetalias{aggarwal_nanograv_2020}, since they use an additional Bayesian ephemeris model, our results are still quite similar. For future data sets with more accurate ephemeris models, we expect these differences to become smaller. Specifically, when using the ephemeris model DE438 in \citet{arzoumanian_nanograv_2020}, the Bayesian ephemeris model, \texttt{BayesEphem}, no longer made a significant difference in common noise parameter estimation.

In the future, given the results in \citet{arzoumanian_nanograv_2020}, in which a detection of a common red-noise process was made, it will be important to include this common process in the signal model for future BWM searches. This additional signal requires introducing two new model parameters. While this would make this method take significantly longer, it may be possible to find improvements in computational costs by using Python vectorization or simply by reducing the resolution of the parameter grid. Preliminary testing shows that a reduction in grid resolution of approximately 20\% still maintains a similar degree of accuracy to the results shown in this work. Even with the addition of two more signal parameters, we expect that this method will still be significantly faster than the traditional MCMC sampling method.

As pulsar timing baselines become longer and PTAs become populated with more pulsars, it will be difficult to use current MCMC sampling techniques to search for GW BWMs, and it will be important to find faster methods to do so. This method is a very efficient way to perform search for BWMs as PTAs continue growing, and data sets become too large for MCMC sampling to be tractable without significant computational resources.

\section*{Acknowledgments}
\begin{acknowledgments}
    The NANOGrav project receives support from National Science Foundation (NSF) Physics Frontiers Center award numbers 1430284 and 2020265.
\end{acknowledgments}

\bibliography{bibliography}{}
\bibliographystyle{aasjournal}



\end{document}